# The infrared imaging spectrograph (IRIS) for TMT: reflective ruled diffraction grating performance testing and discussion

Elliot Meyer*[a,b], Shaojie Chen[a], Shelley A. Wright[a,b], Anna M. Moore[c], James E. Larkin[d], Luc Simard[e], Jêrôme Maire[a], Etsuko Mieda[a,b], Jacob Gordon[b]

[a]Dunlap Institute for Astronomy & Astrophysics, University of Toronto, ON, Canada, M5S 3H4; [b]Department of Astronomy & Astrophysics, University of Toronto, ON, Canada, M5S 3H4; [c]Caltech Optical Observatories, Pasadena, CA, USA, 91125; [d]Division of Astronomy and Astrophysics, University of California, Los Angeles, CA, USA, 90095; [e]Dominion Astrophysical Observatory, National Research Council, Victoria, BC, Canada, V9E 2E7

## ABSTRACT

We present the efficiency of near-infrared reflective ruled diffraction gratings designed for the InfraRed Imaging Spectrograph (IRIS). IRIS is a first light, integral field spectrograph and imager for the Thirty Meter Telescope (TMT) and narrow field infrared adaptive optics system (NFIRAOS). IRIS will operate across the near-infrared encompassing the ZYJHK bands (~0.84 – 2.4μm) with multiple spectral resolutions. We present our experimental setup and analysis of the efficiency of selected reflective diffraction gratings. These measurements are used as a comparison sample against selected candidate Volume Phase Holographic (VPH) gratings (see Chen et al., this conference). We investigate the efficiencies of five ruled gratings designed for IRIS from two separate vendors. Three of the gratings accept a bandpass of 1.19-1.37μm (J band) with ideal spectral resolutions of R=4000 and R=8000, groove densities of 249 and 516 lines/mm, and blaze angles of 9.86° and 20.54° respectively. The other two gratings accept a bandpass of 1.51-1.82μm (H Band) with an ideal spectral resolution of R=4000, groove density of 141 lines/mm, and blaze angle of 9.86°. The fraction of flux in each diffraction mode was compared to both a pure reflection mirror as well as the sum of the flux measured in all observable modes. We measure the efficiencies off blaze angle for all gratings and the efficiencies between the polarization transverse magnetic (TM) and transverse electric (TE) states. The peak reflective efficiencies are $98.90 \pm 3.36\%$ (TM) and $84.99 \pm 2.74\%$ (TM) for the H-band R=4000 and J-band R=4000 respectively. The peak reflective efficiency for the J-band R=8000 grating is $78.78 \pm 2.54\%$ (TE). We find that these ruled gratings do not exhibit a wide dependency on incident angle within ±3°. Our best-manufactured gratings were found to exhibit a dependency on the polarization state of the incident beam with a ~10-20% deviation, consistent with the theoretical efficiency predictions. This work will significantly contribute to the selection of the final grating type and vendor for the IRIS optical system, and are also pertinent to current and future near-infrared astronomical spectrographs.

## 1. INTRODUCTION

The Infrared Imaging Spectrograph[1,2] (IRIS) is an upcoming, first-light instrument for the Thirty Meter Telescope[3] (TMT). It will work in conjunction with NFIRAOS[4], the TMT adaptive optics system. IRIS is a multipurpose instrument that houses a near-infrared imager as well as both a slicer and lenslet integral field spectrographs covering a wavelength range in the near infrared (0.84 – 2.4 μm). IRIS is designed to operate at the diffraction limit of TMT (~ 0.008'' at 1 μm). The spectrograph is designed to operate at a spectral resolution (R) of 4000 and potentially R = 8000 as well. The spectrograph will have a range of field of views of 4.4" x 2.25" to 0.064" x 0.51" dependent on the selected spectral resolution[1].

* meyer@astro.utoronto.ca

This paper covers a large trade study conducted by the IRIS team between potential ruled diffraction gratings (RD) and Volume Phase Holographic[5] (VPH). The goal of the trade study was to determine the optimal grating type for the instrument, either VPH or RD. Traditionally, near-infrared spectrographs have used RD gratings, however optical spectrographs have begun to use VPH gratings[9,10]. VPH gratings offer some advantages over RD gratings such as the grating efficiency being less dependent on polarization and lower scattered light. However, their efficiency suffers greatly from even small incident angle deviations while RD gratings do not exhibit the same issue. Off-blaze performance is crucial for IRIS as the optical design requires many degrees of variation in the incident angle. Presently, the groove densities (< 300 l/mm) required by IRIS to satisfy the wavelength and resolution specifications are difficult to manufacture. It was therefore pertinent that both RD and VPH gratings with the IRIS specifications be tested directly to determine which was the more viable option for this next generation instrument.

To determine the efficiency of grating, a monochromatic light source is often used to determine the amount of flux per diffraction order. There are a number of grating properties that could affect the efficiency, including minor variations in the incident angle, wavefront error across the grating surface, and scattered light properties. Scattered light is an important consideration when designing a spectrograph as the light can be dispersed into other regions of the instrument and can affect the background flux on the detector. For instance, RD gratings that have been manufactured from a used master grating have been found to have significantly more scattered light than uniquely created gratings[6].

In the paper we describe our experimental setup for measuring the peak efficiencies in the surface ruled diffraction gratings. Section 2 covers the IRIS gratings selected for this measurement and their properties. Section 3 gives an overview of the methods and experimental setup. Section 4 covers the experimental errors encountered in your final efficiencies. Section 5 presents the final efficiency and polarization mode results. In Section 6 the comparison between RD and VPH gratings are discussed as well as the overall results of the trade study.

## 2. GRATING SUMMARY

Diffraction ruled gratings follow the traditional grating equation, where the incident beam of light is diffracted into integral diffraction modes depending on the incident angle, wavelength of light, and groove density of the grating,

$$\lambda = \frac{\sigma(\sin\alpha - \sin\beta)}{n} \quad (1)$$

where $\lambda$ is the wavelength of the incident light, $\sigma$ is the groove density in lines/mm, n is the order, and $\alpha$ and $\beta$ are the incident and exit angles respectively. The initial design constraints are that the grating must operate in the 1st order (-1 for the Figures and Tables in this paper) with a detector pixel size of 15 microns (i.e., Hawaii-4RG Teledyne), a pupil diameter of 100 mm and a camera focal length of 370 mm. The current opening angle ($\theta = \alpha + \beta$) of the system is 45 degrees but that can be different depending on the selected components in IRIS. For the lenslet array, an opening angle more than 4° is required which sets an important constraint on which type of gratings can be used in the instrument. Using these parameters along with the grating equation and the dispersion relation (Equation 2), equations for the optimal exit angle and groove density can be derived (Equation 3 and 4). The blaze angle can be found by halving the difference of the incident and optimal exit angle ($\theta_B = (\alpha - \beta)/2$).

$$\frac{d\lambda}{dx} = \frac{\sigma \cos\beta}{n\,f} \quad (2)$$

$$\beta = \tan^{-1}\left[0.4142 - \frac{0.0475}{n}\left(\frac{R}{1000}\right)\right] \quad (3)$$

$$\frac{1}{\sigma} = \frac{81.1}{n\,\lambda}\left(\frac{R}{1000}\right)\cos\beta \quad (4)$$

where f is the focal length of the camera system, and R is the spectral resolution. A summary of the ideal primary grating properties can be seen in Table 1. These equations were used to determine the optimal gratings for each broadband pass (ZYJHK) with the desired spectral resolutions (R=4000 and 8000) for IRIS optical design.

With these grating specifications, we have worked with two vendors to procure ruled diffraction gratings in order to perform efficiency tests for the trade study. We are interested in testing J- and H-band gratings since they require some of the lowest groove densities before K-band for which measurements would require cryogenic cooling. These low groove densities are also harder to manufacture than the high groove density gratings for optical instruments. The gratings tested from Bach Research Corporation[*1] are smaller engineering test gratings whereas the gratings from Changchun Institute of Optics, Fine Mechanics and Physics (CIOMP)[*2] are the science grade gratings with the correct pupil size for IRIS. The gratings have two different coatings (gold and silver) that required two separate mirrors to be procured in order to closely match the grating properties for the efficiency measurements.

CIOMP reported to our team that their J-band, R = 4000 grating had a blaze angle of 9.2° and the H-band, R = 4000 grating had a blaze angle of 9.8° as opposed to the 9.86° that was requested. It is believed that both manufacturers are capable of manufacturing gratings with groove densities at exactly or very near (±1 l/mm) the requested value.

Table 1: Properties of the RD gratings. CIOMP refers to the Changchun Institute of Optics, Fine Mechanics and Physics in Changchun, China. Bach refers to the Bach Research Corporation in Boulder, CO, USA.

| Manufacturer | Spectral Resolution (R) | Bandpass | Blaze Angle (°) | Incident Angle (°) | Groove Density (l / mm) | Physical Size (mm) | Surface Coating |
|---|---|---|---|---|---|---|---|
| CIOMP | 4000 | H-band | 9.86 | 32.36 | 194 | 50x50 | Silver |
| Bach | 4000 | H-band | 9.86 | 32.36 | 194 | 25x25 | Gold |
| CIOMP | 4000 | J-band | 9.86 | 32.36 | 249 | 50x50 | Silver |
| Bach | 4000 | J-band | 9.86 | 32.36 | 249 | 25x25 | Gold |
| Bach | 8000 | J-band | 20.54 | 43.04 | 516 | 25x25 | Gold |

## 3. METHODS AND EXPERIMENTAL SETUP

A similar experiment was performed in our lab for the OSIRIS instrument at Keck Observatory and was recently described by Mieda et al. 2014. The grating efficiencies were measured using a 1.31 μm laser for the J-band gratings and a 1.55 μm laser for the H-band gratings. The lasers are linearly polarized and are fed into a SMF-28 Single Mode Fiber. The fiber was fed into a F240FC-1550 collimator (NA = 0.49) and the output beam through a polarizing filter. The efficiencies were measured in three polarization states: without polarizer (WoP), transverse electric (TE), and transverse magnetic (TM). The TE and TM polarization states arise from the laser being confined to the fiber. The polarizing filter was changed between the TE and TM modes though a 90° rotation. Since the polarization state is determined by the state of the fiber, much care was taken when moving components during the measurement.

The laser was positioned so that the beam would fall on the center of the grating at the required incident angle in Table 1. A Raptor Photonics OWL SWIR 320 camera was used to measure the total flux per diffraction order. The detector is a 320 x 256 pixel InGasAs PIN-Photodiode with a pixel pitch of 30 x 30 μm. The camera was moved to each of the diffraction orders on optical rail and aligned to the optimal position to encompass the entire aperture of the diffraction order on the detector. The radius of the laser spot was approximately 20 pixels. At each order, the detector took 100 frames at a pre-specified exposure time that was determined by ensuring the maximum flux was well below the saturation point of the detector or at the maximum exposure time for very faint high orders to ensure a high signal-to-noise ratio. The entire experiment, except the laser driver, was placed under an aluminum baffle box in order to further reduce the infrared background.

The measurement of the total flux contained in each mode was carried out through an aperture photometry script that utilizes the NASA IDL function 'aper' at its core. A centroid procedure was used to find the ideal central location for the main aperture. The background was subtracted by averaging over a ring aperture outside the main aperture. The total flux was extracted for each frame and the final flux per order was taken as an average of all 100 frames.

[*1] Bach Research Corporation, 2200 Central Avenue Suite D, Boulder, CO 80301, USA
[*2] Changchun Institute of Optics, Fine Mechanics and Physics, Dong Nanhu Road 3888, Changchun, Jilin, 130033, China

The flux values were then used to compute the efficiencies either being compared to the sum of all measured diffraction orders or to the total flux from the pure mirror reflection.

## 4. ERROR ANALYSIS

There are a number of sources of uncertainty that contribute to the overall error in the final efficiency percentage per diffraction order. The primary sources of the uncertainties are the stability of the laser (0.3%), the temperature of the laser (0.3%), incident angle accuracy (1%), camera angle accuracy (2%), and photon noise (approx. 0.1%). The stability of the laser was measured by recording the power over 6 hours[5]. The laser, despite having a thermal electric cooler, required more than a half hour in order to stabilize for measurements. The incident angle could be determined to within 0.2°. Our analysis of the peak efficiency vs angle for the CIOMP J-band, R = 4000 grating determined that within one degree the flux changes only within 1%.

Overall, the camera angle accuracy dominated the measurement errors. This uncertainty arises from the beam not arriving perpendicular to the camera lens. To ensure the camera was as close as possible to a perpendicular orientation the position was varied until the spot had symmetrical distortions on either side of the detector. Still, multiple observations of the same order after moving the camera would result in up to ~2% flux difference.

## 5. RESULTS

### 5.1 Rigorous Coupled Wave Analysis (RCWA)

We performed rigorous couple wave analysis on each of the gratings in order to provide a theoretical expectation for our efficiency results. RCWA is a semi-analytical method used to analyze diffraction by planar gratings using Maxwell's equations in Fourier space. The RCWA scripts allow for our particular gratings to be modeled and theoretical efficiencies for both TE and TM for a particular bandpass to be calculated. In each of the following sections for each grating type a RCWA plot is included.

### 5.2 H-Band, R = 4000 Gratings

Figure 1 shows the theoretical efficiencies calculated using RCWA for a H-band, R = 4000 grating with an updated blaze angle of 9.8° for the CIOMP grating. Figures 2 and 3 present the measured pure reflection efficiencies for the CIOMP H R=4000 grating including WoP, TE, and TM plotted separately as well as ±1° off the bragg angle. The peak reflective efficiencies for the CIOMP and Bach gratings are 98.90 ± 3.36% (TM) and 71.25 ± 3.07% (TE) respectively. The peak all-mode efficiencies are 95.85 ± 6.30% and 71.45 ± 5.05%. Tables 2 and 3 present the reflective efficiency results for the ideal incident angle. Table 4 presents the spatial analysis of the CIOMP grating. The reflective efficiencies were measured for the peak angle in each of the four corners as well as in the center.

Table 2: Reflective efficiencies calculated for the CIOMP H-band, R = 4000 grating at the ideal incident angle of 32.36°

| Order | 1 | 0 | -1 | -2 | -3 | -4 | -5 |
|---|---|---|---|---|---|---|---|
| TE | 10.01 ± 0.32 | 7.57 ± 0.25 | 80.71 ± 2.62 | 2.35 ± 0.08 | 0.52 ± 0.02 | 0.18 ± 0.01 | 0.02 ± 0.001 |
| WoP | 8.60 ± 0.28 | 6.56 ± 0.21 | 85.13 ± 2.76 | 2.18 ± 0.07 | 0.49 ± 0.02 | 0.19 ± 0.01 | 0.05 ± 0.002 |
| TM | 0.91 ± 0.03 | 1.32 ± 0.04 | 98.90 ± 3.36 | 1.23 ± 0.04 | 0.32 ± 0.01 | 0.28 ± 0.01 | 0.21 ± 0.01 |

Table 3: Reflective efficiencies calculated for the Bach H-band, R = 4000 grating at the ideal incident angle of 32.36°

| Order | 1 | 0 | -1 | -2 | -3 | -4 | -5 |
|---|---|---|---|---|---|---|---|
| TE | 14.54 ± 0.54 | 4.85 ± 0.19 | 71.25 ± 3.07 | 7.98 ± 0.34 | 0.81 ± 0.03 | 0.18 ± 0.01 | 0.04 ± 0.001 |
| WoP | 13.74 ± 0.48 | 4.45 ± 0.16 | 71.79 ± 2.81 | 8.80 ± 0.31 | 1.12 ± 0.04 | 0.38 ± 0.01 | 0.19 ± 0.01 |
| TM | 6.65 ± 0.24 | 1.27 ± 0.04 | 70.78 ± 2.68 | 13.84 ± 0.48 | 3.61 ± 0.13 | 1.85 ± 0.06 | 1.45 ± 0.05 |

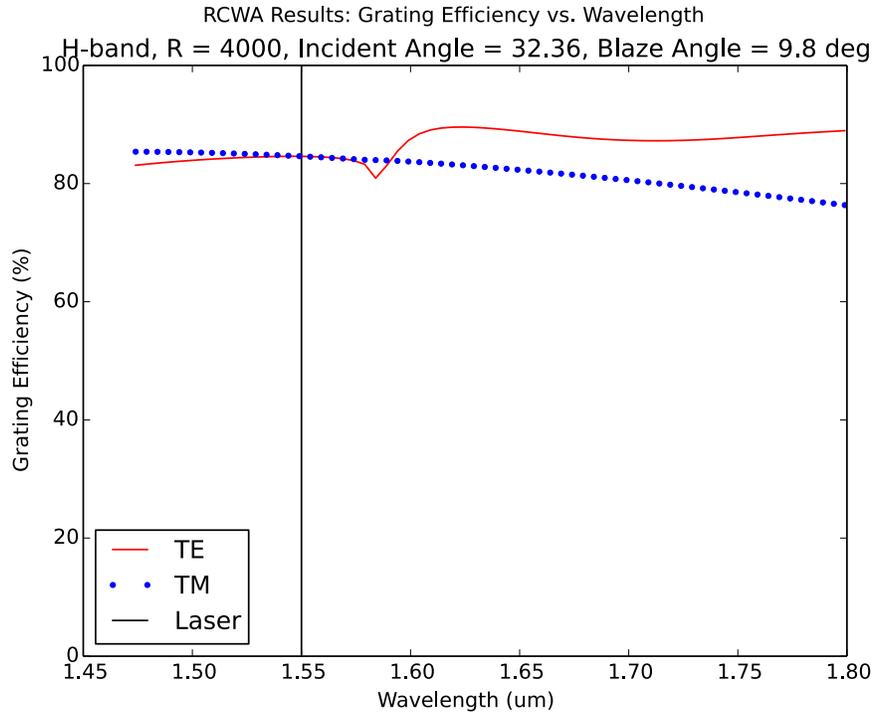

Figure 1: The RCWA results for a H-band grating with a blaze angle of 9.8°, sigma of 194 l/mm, R = 4000, and an incident angle of 32.36° in the -1 order.

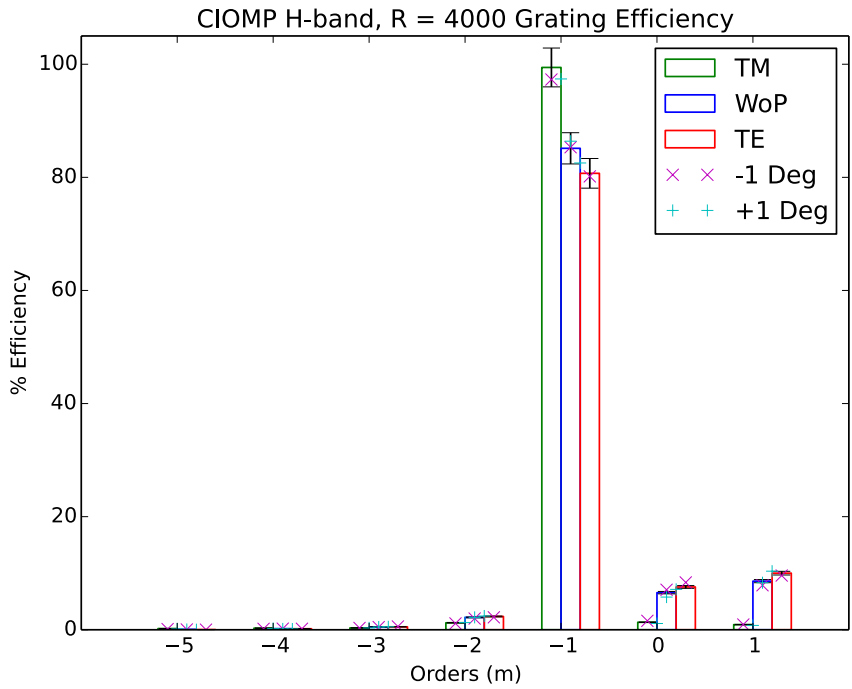

Figure 2: The reflective efficiency results for the CIOMP H-band, R = 4000 grating. The +/- 1° efficiency measurements are marked with x and + symbols.

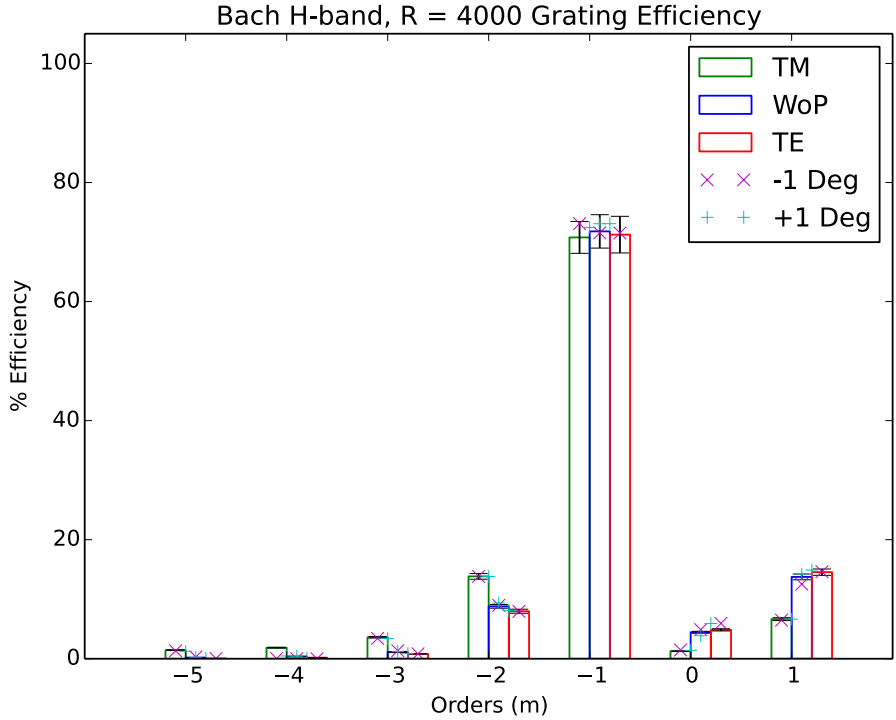

Figure 3: The reflective efficiency results for the Bach H-band, R = 4000 grating. The +/- 1° efficiency measurements are marked with x and + symbols.

Table 4: The spatial grating efficiency for the CIOMP H-band, R = 4000 grating. The measurements were made without the polarizer. The measurements were made in the four corners of the grating as well as the center.

| | | |
|---|---|---|
| 86.74 ± 2.90% | | 84.55 ± 2.84% |
| | 85.13 ± 2.76% | |
| 80.78 ± 2.72% | | 85.46 ± 2.86% |

## 5.3 J-Band, R = 4000 Gratings

Figure 4 shows the theoretical efficiencies calculated using RCWA for a J-band, R = 4000 grating with an updated blaze angle of 9.2° for the CIOMP grating. Figures 5 and 6 present the reflection efficiencies. Figure 7 exhibits the peak efficiency vs incident angle for all polarization states to a maximum of 3° from the ideal incident angle. There are no WoP measurements for the Bach grating due to time constraints for our measurements. Regardless, the WoP value should lie somewhere between the TE and TM mode efficiencies depending on the polarization state during the measurement. The peak reflective efficiencies for these gratings are 84.99 ± 2.74% (TM) and 67.98 ± 2.19 (TM) for CIOMP and Bach respectively. Tables 4 and 5 present the reflective efficiency results for the ideal incident angle. The peak all-mode efficiencies are 83.80 ± 5.05% and 66.74 ± 4.43%.

Table 4: Reflective efficiencies calculated for the CIOMP J-band, R = 4000 grating at the ideal incident angle of 32.36°

| Order | 1 | 0 | -1 | -2 | -3 | -4 |
|---|---|---|---|---|---|---|
| TE | 9.96 ± 0.32 | 21.30 ± 0.69 | 71.76 ± 2.31 | 0.04 ± 0.001 | 0.09 ± 0.003 | |
| WoP | 4.94 ± 0.16 | 11.07 ± 0.36 | 84.87 ± 2.74 | 1.45 ± 0.04 | 0.44 ± 0.01 | 0.53 ± 0.02 |
| TM | 4.38 ± 0.14 | 9.81 ± 0.32 | 84.99 ± 2.74 | 1.24 ± 0.04 | 0.46 ± 0.01 | 0.53 ± 0.02 |

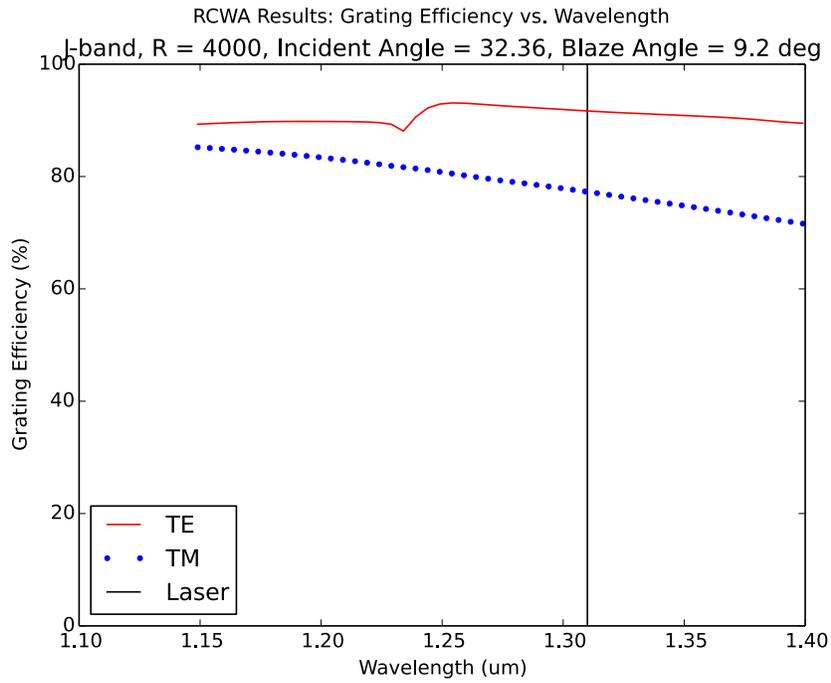

Figure 4: The RCWA results for a J-band grating with a blaze angle of 9.2°, sigma of 249 l/mm, R = 4000, and an incident angle of 32.36° in the -1 order.

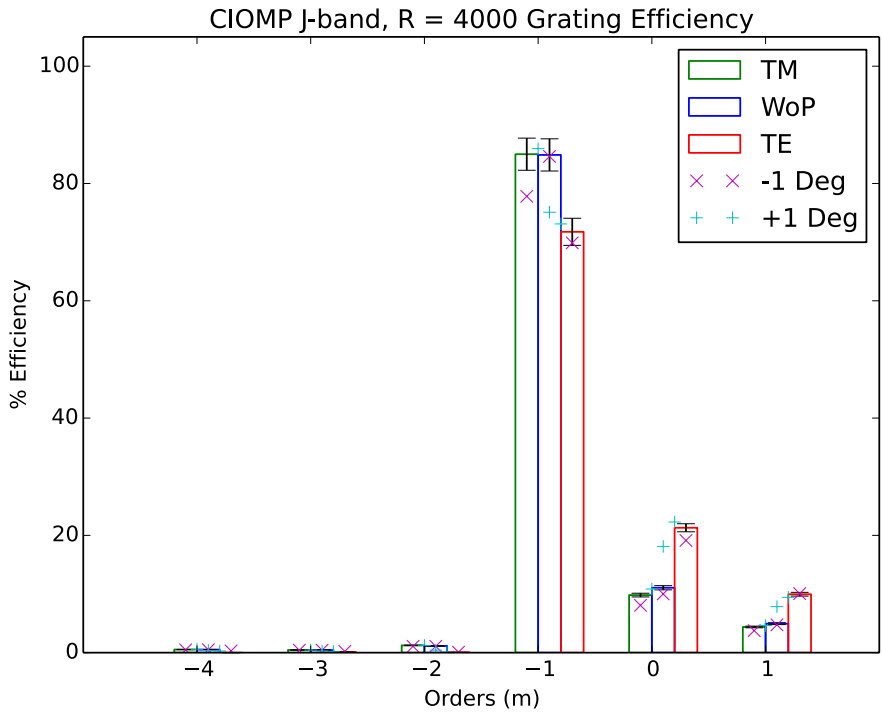

Figure 5: The reflective efficiency results for the CIOMP J-band, R = 4000 grating. The +/- 1° efficiency measurements are marked with x and + symbols.

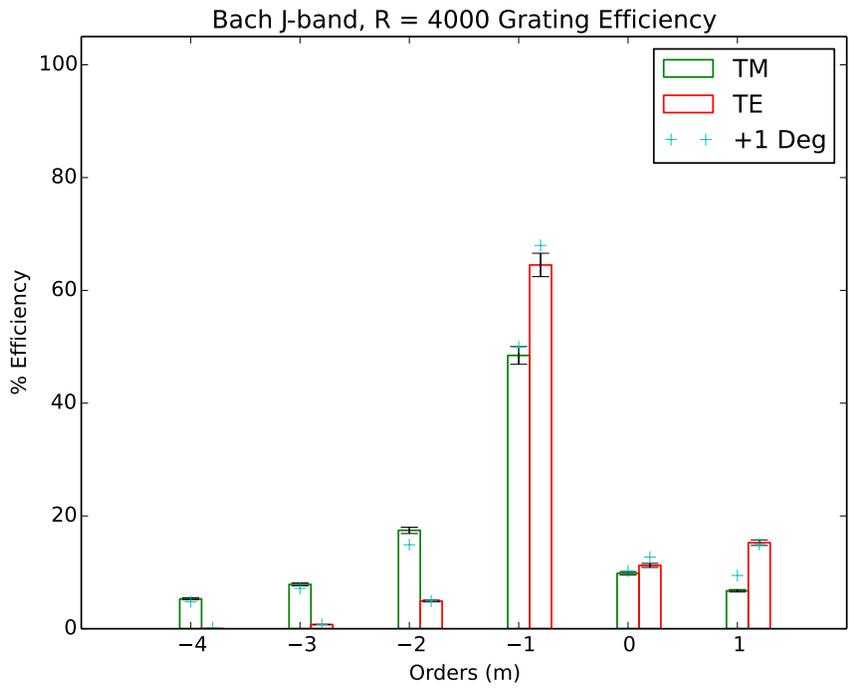

Figure 6: The reflective efficiency results for the Bach J-band, R = 4000 grating. The +/- 1° efficiency measurements are marked with x and + symbols.

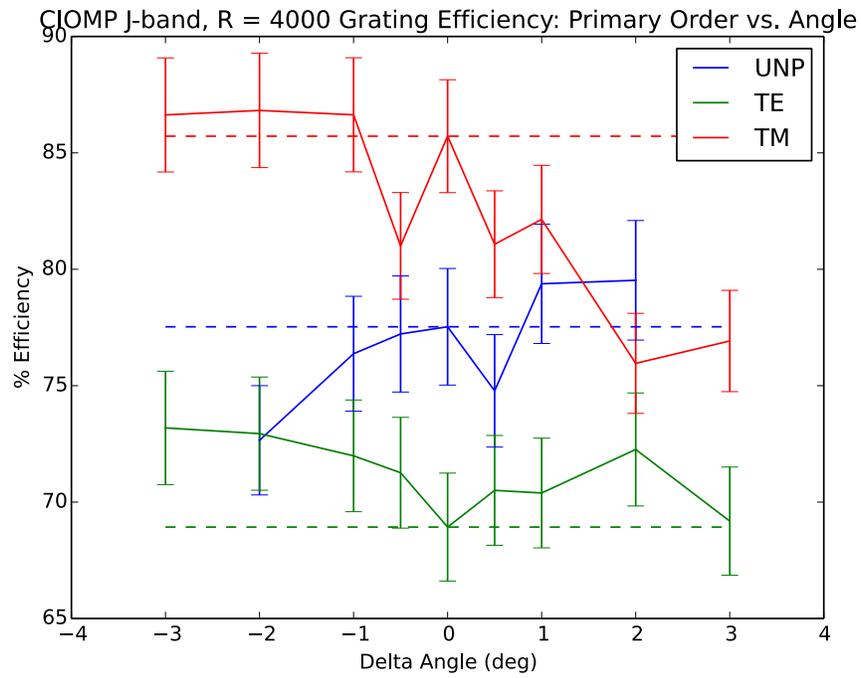

Figure 7: Measurements of the efficiency of the primary order vs. incident angle deviation. The error bars represent the one sigma errors. The dotted lines are positioned at the ideal incident angle value in order to judge how greatly the efficiency changes.

Table 5: Reflective efficiencies calculated for the Bach J-band, R = 4000 grating at the ideal incident angle of 32.36°

| Order | 1 | 0 | -1 | -2 | -3 | -4 |
|---|---|---|---|---|---|---|
| TE | 15.26 ± 0.49 | 11.23 ± 0.36 | 64.50 ± 2.08 | 4.9 ± 0.16 | 0.73 ± 0.02 | |
| TM | 6.73 ± 0.22 | 9.82 ± 0.32 | 48.46 ± 1.56 | 17.44 ± 0.56 | 7.86 ± 0.25 | 5.30 ± 0.17 |

## 5.4 J-Band, R = 8000 Grating

Figure 8 shows the theoretical efficiencies calculated using RCWA for a J-band, R = 8000 grating. Figure 9 presents the measured reflection efficiencies. Only the primary order and $0^{th}$ order could be measured along with ±1° off the bragg angle for those orders as well as the -2 order. This was due to both time constraints as well as difficultly in measuring the large angles of the R = 8000 orders. Since the efficiencies are not known for all orders there is no 'all-modes' result. The peak reflective efficiency for this grating was 78.78 ± 2.54% (TE). Table 6 presents the reflective efficiency results for the bragg angle.

Table 6: Reflective efficiencies calculated for the Bach J-band, R = 8000 grating at the ideal incident angle of 43.04°.

| Order | 0 | -1 |
|---|---|---|
| TE | 22.30 ± 0.71 | 78.78 ± 2.54 |
| TM | 7.13 ± 0.23 | 75.18 ± 2.42 |

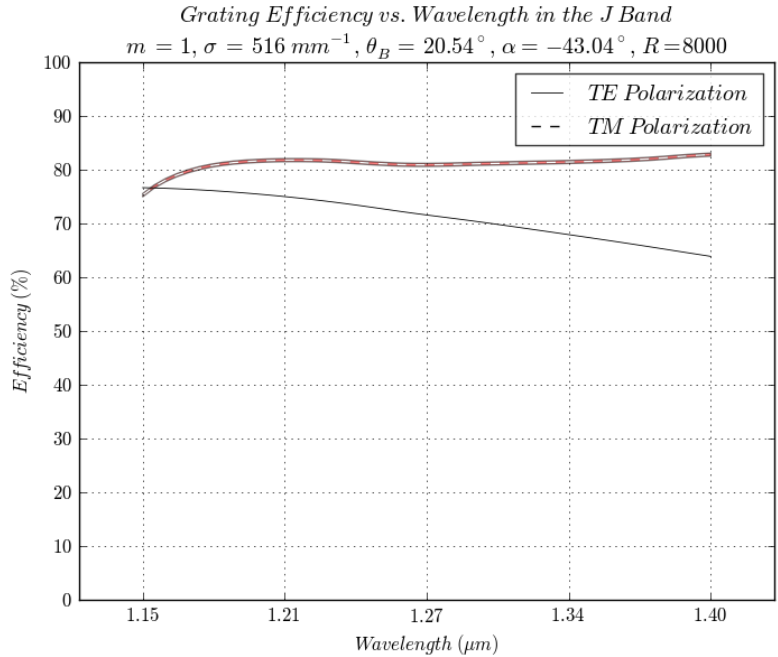

Figure 8: The RCWA results for a J-band grating with a blaze angle of 20.54°, sigma of 516 l/mm, R = 8000, and an incident angle of 32.35° in the -1 order.

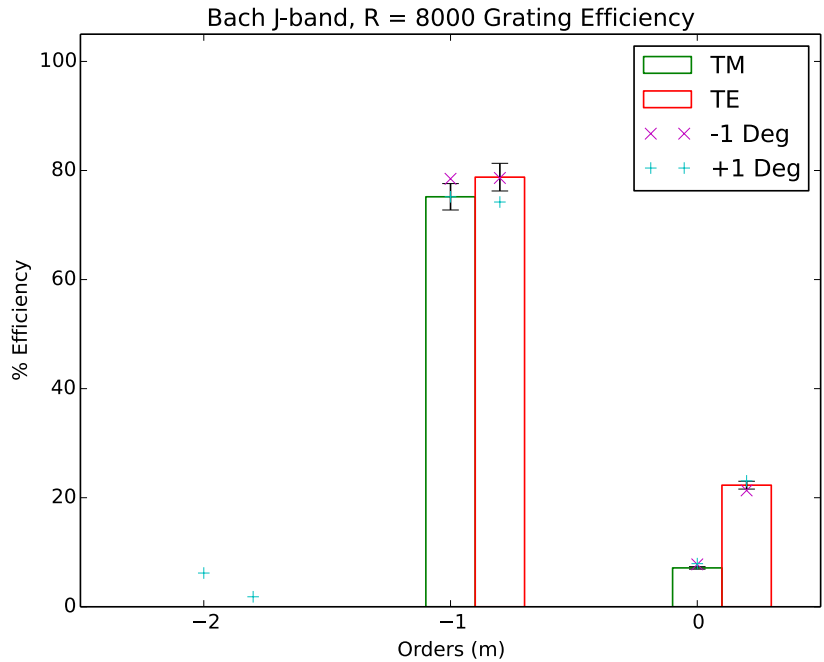

Figure 9: The reflective efficiency results for the Bach J-band, R = 8000 grating. The +/- 1° efficiency measurements are marked with x and + symbols.

## 6. DISCUSSION

In summary, the CIOMP gratings had peak reflective efficiencies of 98.90 ± 3.36 (TM) and 84.99 ± 2.75% (TM) for H- and J-band R = 4000 and the Bach gratings had 71.25 ± 3.07 (TE), 64.50 ± 2.08% (TE), 78.78 ± 2.54% (TE) for the H- and J-band R = 4000, and J-band R = 8000 gratings respectively. All peak order errors lie within 2-3.5%. It is clear that based on efficiencies of the gratings that were measured that the CIOMP gratings are the closest to the theoretical RCWA derived values. The Bach J-band, R = 8000 grating peak efficiency is very close to the predicted efficiency through RCWA analysis. The 'all-mode' efficiencies have 2-3 times greater error than the reflective efficiencies due to the dependency on a greater number of measurements, which is why we prefer the reflective efficiencies for our analysis. Most of the reflective efficiencies are contained within the errors of the 'all-mode' efficiencies, making drawing any conclusions at the present time regarding scattered light difficult. In addition, the light is primarily constrained within the primary order and the $0^{th}$ order for the CIOMP gratings whereas the light is more dispersed among all orders for the Bach gratings. This is an important grating property for if the light is concentrated in only a few orders it will be easier to remove from the instrument.

The 98.90 ± 3.36% efficiency result for the CIOMP H-band, R = 4000 grating was re-measured 3 times and the same results, within 1σ, were obtained. Since the measured value is significantly greater than the RCWA prediction, we believe that the grating may have different properties other than the minor blaze angle difference that was reported by CIOMP through private correspondence. The Bach Research Corp did not perform any tests on their gratings to determine the true values, however they mentioned in private correspondence that their manufacturing could be improved to account for the lower efficiencies that we observed. In addition, our investigation of the H-band, R = 4000 gratings used a monochromatic laser and so it is possible that the grating has particularly high efficiency at 1.55 μm. The sum of all TM mode reflective efficiencies is still, within 1σ, 100%. It is also worth noting that the RCWA calculations predict a negligible difference between TE and TM efficiencies at 1.55 μm yet we observe almost a 20% difference.

The primary goal of this trade study was to compare the properties of RD and VPH gratings in the near infrared with relatively low groove densities (200 l/mm). The RD gratings are much better than the VPH gratings when the peak efficiency is compared to a variation in the incident angle. The RCWA analysis outlines the efficiency dependence on the incident angle. The measured CIOMP J-band, R = 4000 grating efficiency vs. incident angle is presented in Figure 11. Within each polarization state the efficiency decreases by less than 10% when off the bragg angle by 3°. In contrast, the VPH gratings can exhibit a decrease in efficiency of greater than 20% when the incident angle is similarly changed[5]. This is very important to the design of IRIS as the gratings will be required to accept an incident angle difference of up to 4.4°[8]. The VPH gratings will cause a large drop in efficiency over the spatial field-of-view, whereas the effect will be much reduced if RD gratings are used.

Gratings from both manufacturers exhibited a large dependence on polarization state. The difference between TM and TE state efficiencies were greater than 10%. The large differences in efficiency with polarization are not ubiquitous, however the RCWA analysis predicts a wide variability with wavelength. On the contrary, the VPH gratings exhibit a lower dependence of the efficiencies on polarization than the RD gratings[5].

The spatial efficiencies for the CIOMP H-band, R = 4000 grating were measured. Within 1$\sigma$, the majority of the grating has an efficiency of 85%. However, the lower left corner has a lower efficiency, which is most likely caused by a higher wavefront error in this region. This is important as the entire grating area will be illuminated in the instrument and the majority of the grating has equal efficiencies within our measured error. We wish to do a similar analysis for the other RD gratings as well as the VPH gratings.

In conclusion, we have shown that current RD manufacturers are able to make high efficiency near-infrared, low groove density gratings. The efficiencies of these RD compare favorably to those measured for VPH gratings. These gratings will be of great use for future instruments for Extremely Large Telescopes as well as space-based observatories.

## 7. ACKNOWLEDGEMENTS

The authors would like to thank the Changchun Institute of Optics, Fine Mechanics and Physics and the Bach Research Institute for their support and cooperation throughout this study. We thank the support and resources offered by the Dunlap Institute of Astronomy & Astrophysics at University of Toronto. The Dunlap Institute is funded through an endowment established by the David Dunlap family and the University of Toronto. The authors gratefully acknowledge the support of the TMT partner institutions. They are the Association of Canadian Universities for Research in Astronomy, California Institute of Technology, Department of Science and Technology India, National Astronomical Observatories of the Chinese Academy of Science, the National Astronomical Observatory of Japan, and the University of California. The TMT project is planning to build the telescope facilities on Mauna Kea, Hawaii. The authors wish to recognize the significant cultural role and reverence that the summit of Mauna Kea has always had with the indigenous Hawaiian community.

## REFERENCES

[1] Larkin, J. E., Moore, A. M., Barton, E. J., Bauman, B. J., Bui, K., Canfield, J. M., Crampton, D., Delacroix, A., Fletcher, J. M., Hale, D., Loop, D. J., Niehaus, C. N., Phillips, A. C., Reshetov, V. A., Simard, L., Smith, R. M., Suzuki, R., Usada, T., Wright, S. A., "The infrared imaging spectrograph (IRIS) for TMT; instrument overview", Proc. SPIE 7735, 7735-79 (2010)
[2] Moore, A.M., Larkin, K. E., Wright, S. A., Bauman, B., Dunn, J., Ellerbroek, B., Phillips, A. C., Simard, L., Suzuki, R., Zhang, et al. "The infrared imaging spectrograph (IRIS) for TMT: instrument overview", Proc. SPIE 9147-76 (2014)
[3] Sanders, G. H., "The Thirty Meter Telescope (TMT): An International Observatory, " Journal of Astrophysics and Astronomy 34, 81-86 (2013)
[4] Herriot, G., Anderson, D., Atwood, J., Byrnes, P., Boucher, M-A., Boyer, C., Caputa, K., Correia, C., Dunn, J., Ellerbroek, B., Fitzsimmons, J., Gilles, L., Hickson, P., Hill, A., Kerley, D., Pazder, J., Reshetov, V., Roberts, S., Smith, M., Véran, J-P., Wang, L., Wevers, I. "TMT NFIRAOS: adaptive optics system for the Thirty Meter Telescope", Proc. SPIE (2012)
[5] Chen, S., Meyer, E., Wright, S. A., Moore, A. M., Larkin, J. E., Marie, J., Mieda, E., Simard, L. "The infrared imaging spectrograph (IRIS) for TMT: volume phase holographic grating performance testing and discussion", Proc. SPIE 9147-334 (2014)

[6] Woods, T. N., Wrigley, R. T. III, Rottman, G. J., Haring, R. E., "Scattered-light properties of diffraction gratings," Applied Optics 33, 4273-4285 (1994)
[7] Mieda, E., Wright, S. A., Larkin, J. E., Graham, J. R., Adkins, S. M., Lyke, J. E., Campbell, R. D., Maire, J., Do, T., Gordon, J., "Efficiency Measurements and Installation of a New Grating for the OSIRIS Spectrograph at Keck Observatory", Publications of the Astronomical Society of the Pacific 126, 937, 250-263 (2014)
[8] Moore, A. M., Bauman, B. J., Barton, E. J., Crampton, D., Delacroix, A., Larkin, J. E., Simard, L., and Wright, S. A., "The Infrared Imaging Spectrograph (IRIS) for TMT: Spectrograph Design", Proc. SPIE, 7735-87 (2010)
[9] Hou, Y., Xhu, Y., Hu, Z., Wang, L., Wang, J., "Performance and Sensitivity of Low Resolution Spectrograph for LAMOST", Proc. SPIE, 7735 (2010)
[10] Bernstein, G. M., Athey, A. E., Bernstein, R., Gunnels, S. M., Richstone, D. O., Shectman, S. A. "A volume-phase holographic spectrograph for the Magellan telescopes" Proc. SPIE, 4485 (2002)